\documentclass[aps,prc,preprint,groupedaddress]{revtex4}
\usepackage[dvipdfmx]{graphicx}
\usepackage{ulem}

\newcommand{\appropto}{\mathrel{\vcenter{
  \offinterlineskip\halign{\hfil$##$\cr
    \propto\cr\noalign{\kern2pt}\sim\cr\noalign{\kern-2pt}}}}}

\begin{document}


\title{Dependence of  Weak Interaction Rates on  the Nuclear Composition during Stellar Core Collapse}


\author{Shun Furusawa}
\email{furusawa@fias.uni-frankfurt.de}
\affiliation{Frankfurt Institute for Advanced Studies, J.W. Goethe University, 60438 Frankfurt am Main, Germany}
\author{Hiroki Nagakura}
\affiliation{TAPIR, Walter Burke Institute for Theoretical Physics, Mailcode 350-17, California Institute of Technology, Pasadena, CA 91125, USA}
\author{Kohsuke Sumiyoshi}
\affiliation{Numazu College of Technology, Ooka 3600, Numazu, Shizuoka 410-8501, Japan}
\author{Chinami Kato}
\affiliation{Advanced Research Institute for Science and Engineering, Waseda University, 3-4-1 Okubo, Shinjuku, Tokyo 169-8555, Japan}
\author{Shoichi Yamada}
\affiliation{Advanced Research Institute for Science and Engineering, Waseda University, 3-4-1 Okubo, Shinjuku, Tokyo 169-8555, Japan}

\date{\today}

\begin{abstract}
We investigate the influences of the nuclear composition 
  on the weak interaction rates of heavy nuclei during the  core collapse of massive stars.
The nuclear abundances 
in nuclear statistical equilibrium (NSE)
are calculated by some  equation of state (EOS) models
including  
 in-medium effects on nuclear masses.
We systematically examine 
the sensitivities of 
electron capture and neutrino-nucleus scattering on heavy nuclei 
to  the nuclear shell effects and  the single nucleus approximation. 
We find that 
 the washout of shell effects at high temperatures brings significant change to weak rates
by  smoothing the nuclear abundance distribution: 
the electron capture rate decreases by $\sim$20$\%$ in the early phase   and
 increases by $\sim$40$\%$  in the late phase  at most,
while the cross section for neutrino-nucleus scattering is reduced by  $\sim$ $15\%$.
This is because the open-shell nuclei  become abundant  instead of  those with closed neutron shells as the shell effects disappear.
We also find that the single-nucleus description based on the average values leads to 
underestimations of weak rates. Electron captures  and neutrino coherent scattering on heavy nuclei  are  reduced by $\sim$80$\%$  in  the early phase 
and by $\sim$5$\%$ in the late phase, respectively.
These results indicate that  NSE like EOS accounting for shell washout is indispensable for the reliable  estimation of weak  interaction rates  in simulations of core-collapse supernovae. 
\end{abstract}

\maketitle

\section{Introduction \label{intro}}
Reliable nuclear physics as input  data  is indispensable for simulations of  core collapse supernovae
which  are considered to  occur at the end of the evolution of massive stars
 and lead to the emissions of neutrinos and  gravitational waves,  the synthesis  of heavy elements and  the formations of  a neutron star  or a black hole \cite{janka12,kotake12,burrows13}.
One of the underlying problems in these events is the uncertainties in input data for numerical simulations such as  the equation of state (EOS) and  weak interaction rate for hot and dense matter. 
Supernova matter is composed of nucleons, nuclei, electrons, photons and neutrinos 
 and  its temperature is high enough to achieve chemical equilibrium for all strong and electromagnetic reactions.
The  composition of nuclear matter 
 is determined as a function of  temperature $T$, density $\rho_B$ and electron fraction $Y_e$
 by EOS models. 
The weak interactions on the nuclear components
 play important roles in  the dynamics of core collapse supernovae through 
the evolution of the lepton fraction 
 \citep{hix03,lentz12}. 

 Nuclear composition depends on the models employed for 
nuclei and nuclear matter 
and is affected by uncertainties. 
These equally come from finite density and temperature effects 
\citep{aymard14,agrawal14a,steiner13,buyukcizmeci13} 
and neutron/proton composition not yet produced and studied in terrestrial laboratories.
Roughly speaking, there are two types of EOS models for supernova simulations.
The single nucleus approximation (SNA),  widely used over the past 2
decades, considers that the ensemble of heavy nuclei may be
represented by a single nucleus  \cite{lattimer91,shen98a,shen98b,shen11}.
As a consequence, it can not
account for the whole nuclear composition, which is indispensable
for an accurate estimation of weak interaction rates.
Furthermore,  even the average mass and proton numbers and mass fraction of heavy nuclei may not be reproduced correctly by the representative nucleus 
 \cite{burrows84,furusawa17c}.
The other type of EOS model is the multi-nucleus EOS,
in which the full ensemble of nuclei is solved and
 nuclear statistical equilibrium (NSE) abundance is obtained
for each set of thermodynamical conditions \cite{botvina04, botvina10,buyukcizmeci14,hempel10,sheng11}.

We constructed an multi-nucleus EOS including various in-medium effects such as formation of  nuclear pastas and washout of shell effects of heavy nuclei \cite{furusawa11,furusawa13a,furusawa17a}.
The shell effects are derived from the structure of nuclei in  the ground state  and smeared out  completely at $T \sim$ 2.0~-~3.0~MeV \citep{brack74, bohr87, sandulescu97,nishimura14}.
This washout  effect changes the nuclear component  considerably \cite{furusawa17a},
whereas it has not been taken into account in all supernova EOS models.
In most EOS models \cite{lattimer91,shen98a,shen98b,shen11, botvina04, botvina10,buyukcizmeci14},  shell effects are neglected in the first place  
and set to be completely smeared out  even at zero temperature.
Other  EOS models  assume full shell effects at any temperature \cite{hempel10,sheng11}.
Some works, however, show that the  evaluation of shell energies makes a large difference in   the nuclear composition \cite{furusawa11,buyukcizmeci13}.
Recently,  Raduta et al. reported that the shell  quenching, which is a drop in the shell effect for neutron rich nuclei, affects 
up to $\sim$30$\%$ of  the average electron capture rate during  the core collapse 
 owing to the modification of NSE abundance \cite{raduta16}.

Neutrino scattering and electron capture on heavy nuclei
are the main weak interactions in collapsing  cores  to determine the evolution of lepton fraction and the size of bounce cores.
Those reaction rates  at given ($\rho_B,T,Y_e$) are obtained by 
folding the NSE abundance with the individual rates of all nuclei.
When the single-nucleus EOS is utilized to calculate them,  we have no choice but to substitute the representative nucleus for  the ensemble of heavy nuclei. 
 Even with  the multi-nucleus EOS, 
 we often estimate the weak rates
by using the average values, such as average mass and proton numbers  of nuclear ensemble, which are generally listed in EOS data tables for simulations. 
These prescriptions in the single nucleus descriptions may bear artificial errors from the complete folding results.

The calculation of  electron captures contains more ambiguities than that of  neutrino-nucleus scattering and
 provides a large uncertainty in supernova simulations. 
To obtain the total electron capture rates, 
we need the rates of individual nuclei. 
The first tabulated data of  these,  
based on shell model calculations, was provided by Fuller et al. (FFN \cite{fuller82}).
Oda et al. (ODA \cite{oda94}) and Langanke $\&$ Mart$\rm{\acute{\i}}$nez-Pinedo
(LMP \cite{langanke00}) constructed tabulated data for sd- and pf-shell
 nuclei, respectively, based on shell model calculations with effective interactions and experimental energy levels.
Langanke et al. (LMSH \cite{langanke03})  employed the Monte Carlo approach 
with a random
phase approximation (RPA)  for heavier nuclei in the pfg/sdg-shell.
Unfortunately, these calculations for  electron capture rates were aimed at  nuclei along the $\beta-$stable line,
whereas collapsing cores encounter the neutron-rich nuclei such as ~$^{78}$Ni,
 whose rates are yet to be developed from sophisticated models 
 \cite{sullivan16}.
In general,  we adopt any approximation formula 
for the nuclei with no data available.

For decades,  the  approximation formula for electron-captures provided by  Bruenn  \cite{bruenn85} 
has been widely utilized, 
it has been pointed out that the formula underestimates the rates at finite temperature,
  since  they ignored the thermal excitation of neutrons in the daughter nucleus and cut off the reactions for
 the nuclei with the neutron numbers larger than $40$ \cite{langanke03}.
It is known that the adoption of  the Bruenn's rate for the representative nucleus of SNA
results in the overestimations of the lepton fraction and mass of bounce core
and, hence, we should utilize 
more sophisticated data or formula with the NSE like EOS \cite{hix03,langanke03,lentz12}. 
Juodagalvis et al. \cite{juodagalvis10} 
calculated the electron capture rates of  nuclei more than 2200 based on the same RPA technique for LMSH with a Fermi – Dirac parametrization  and constructed the data averaged over  NSE abundance, 
although the individual rates  were not released.
We used their data  in supernova simulations \cite{nagakura16},  
but the nuclear  composition may be inconsistent between  the NSE calculation used in the preparation for their data and 
our  EOS  model that includes in-medium effects and was adopted in the simulations \cite{furusawa13a}.

The purpose of this study is to clarify the impact of the uncertainties in the nuclear  composition provided by EOS models  on weak interaction rates. 
We focus on the shell effects of heavy nuclei,
 especially on the  washout of them.
We also report the deviation of the approximate weak rates 
 for the average values and the most probable nucleus in the single nucleus descriptions
from
 the accurate rates  obtained with NSE abundances and individual rates. 

This article is organized as follows.
The EOS models and the nuclear compositions realized in the collapsing core are described in section~\ref{sec:eos}.
 In section~\ref{sec:weak}, we discuss the weak interaction rates based on them 
with emphases  on the impact of shell effects and  the errors of SNA. 
The paper is wrapped up with a summary and some discussions in section~\ref{sec:sum}.

\section{Equation of state and nuclear composition \label{sec:eos}}
\begin{figure}
\includegraphics[width=12cm]{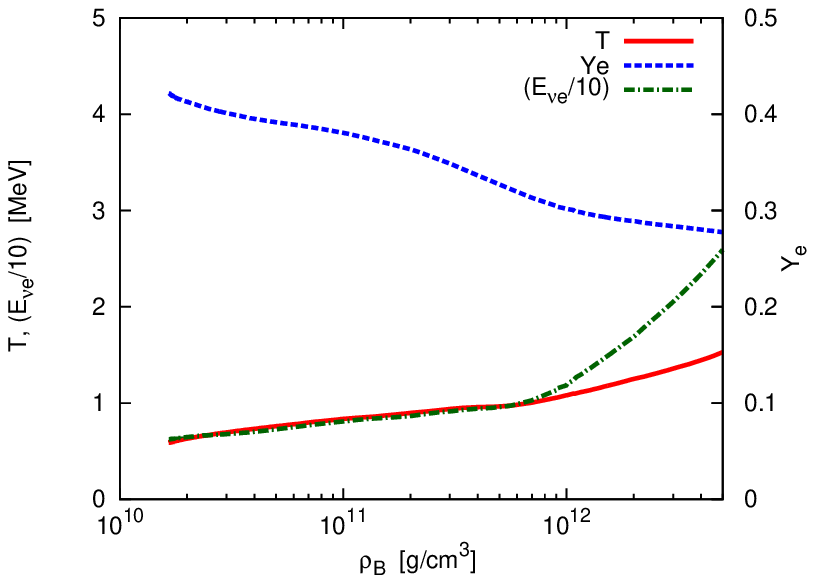}
\caption{Temperature (red solid line), electron fraction (blue dashed line) and average energy of electron-type neutrinos (green dashed-dotted line) as a function of density at the center of the collapsing core  in the reference   supernova simulation  of 11.2~${\rm M_{\odot}}$ progenitor \cite{nagakura16}. }
\label{fig_tye}
\end{figure}

We calculate  the nuclear abundance in collapsing cores based on  the  EOS models,
 the details of which are given in Furusawa et al.  \cite{furusawa11,furusawa13a,furusawa17a}.
The thermodynamical states are given by $T$,  $\rho_B$ and $Y_e$
and those values in the center of  the collapsing core are
 taken from a recent supernova simulation \cite{nagakura16} with the progenitor of 11.2~${\rm M_{\odot}}$ \cite{woosley02}.
In  the simulation,  
the  Boltzmann neutrino-radiation hydrodynamics is exactly solved in spherical symmetry.
We utilize the previous version of our EOS \cite{furusawa13a}
and  the data of electron capture rates provided by Juodagalvis et al. \cite{juodagalvis10}.
Figure~\ref{fig_tye} shows the evolutions of  $T$ and  $Y_e$ as functions of $\rho_B$  at the center of  the collapsing core before the core bounce. 
The simulation starts at $\rho_B \sim 2 \times 10^{10}$ ~g/cm$^{3}$ and  the core bounce occurs when the central density reaches $\rho_B \sim 3 \times  10^{14}$ ~g/cm$^{3}$. 

The model free energy density of our EOS reproduces the ordinary NSE results at low densities and temperatures  and makes a continuous transition to the supra-nuclear density EOS.
The thermodynamical quantities and nuclear abundances  as functions of $\rho_B$, $T$ and $Y_e$ are obtained by minimizing the model free energy with 
respect to the number densities of nuclei and nucleons under the baryon and charge  conservations. 
The model free energy is expressed as
\begin{eqnarray}
\ f = f_{p,n}+\sum_{j}{n_j (E^t_{j} + M_{j}) } + \sum_i {n_i (E^t_{i} + M_{i})}, 
\end{eqnarray}
where $n_{j/i}$ is the number density of the  individual nucleus, index $j$ specifying  a light nucleus with the proton number $Z_j\leq 5$ and
 index $i$ meaning a heavy nucleus with  $6 \leq Z_i \leq 1000$.
The free energy density of the nucleon vapor outside nuclei $f_{p,n}$ is 
 calculated by the  relativistic mean field (RMF) with the TM1 parameter set \cite{sugahara94}.
The translational energies of heavy and light nuclei $E^t_{i/j}$ are based on 
 that for the ideal Boltzmann gas with the same excluded-volume effect as that in  Lattimer's EOS \cite{lattimer91}.
The masses of light nuclei, $M_j$, are evaluated by a quantum approach, in which the self- and Pauli-energy shifts are taken into account \cite{typel10,roepke09}. 
The masses of heavy nuclei are  assumed to be the sum of  bulk, Coulomb, surface  and shell energies: $M_i=E_i^B+E_i^C+E_i^{Su} +E_i^{Sh}$.
The bulk energies  are obtained  via
the same RMF for free nucleons,
 while various modifications at finite density and temperature are accounted for in evaluations of  surface and Coulomb energies. 

  The shell energies
at zero temperature, $E_{i0}^{Sh}$, 
  are  obtained
 from the experimental or theoretical mass data \cite{audi12,koura05} by subtracting our liquid drop mass formula, which does not include the shell effects, $(M_i^{LDM}=E_i^B+E_i^C+E_i^{Su})$ in the vacuum limit as $E_{i0}^{Sh} =M_i^{data} -[ M_i^{LDM} ]_{vacuum}$.
We take the washout of the shell effect into account approximately as follows: 
\begin{eqnarray}\label{eq:sh}
 E_{i}^{Sh}(T)= 
E_{i0}^{Sh} \displaystyle{\frac{\tau}{{\rm sinh}\tau}} .  
\end{eqnarray} 
The  factor $\tau/{\rm sinh}\tau$ is derived   by the analytical study for the  single particle motion of nucleon outside the closed shell \citep{bohr87}. %
The normalized factor $\tau$ is defined as $\tau=2 \pi^2 T/\epsilon_{sh}$
with the energy spacing of  the shells, $\epsilon_{sh} = 41 A_i^{-1/3}$~MeV,  where $A_i$ is  the mass number of nucleus~$i$. 
This formulation  can reproduce  the feature of  washout in that the shell energies disappear around $T\sim$~2.0－3.0 MeV. 
Note that we  ignore the density dependence of shell effects, which is considered in the original EOS models \cite{furusawa11,furusawa17a},
 for simplicity and since it is negligible in this study.

 To clarify the impact of the shell washout  on weak  rates,
we prepare  Models FS (full shells), WS (washout shells) and NS (no shells). 
Model FS ignores the washout, 
dropping the factor $\tau/{\rm sinh}\tau$ in Eq.~(\ref{eq:sh}). 
They may be regarded as a surrogate for  the   ordinary NSE calculations \cite{timmes99,hempel10}, which also neglect the washout effect,
 but are widely utilized  to estimate electron capture rates in supernova matter \cite{juodagalvis10,sullivan16}.
The previous version of our EOS used in the reference supernova simulation also lacks  this effect  \cite{furusawa13a,nagakura16}. 
Model WS is the new EOS \cite{furusawa17a}, in which we take it. 
We also prepare Model NS for comparison, in which the
shell effects themselves are not considered at all  ($E_{i0}^{Sh}=0$).
This model is similar to the EOS models \cite{lattimer91,shen11,buyukcizmeci13}, in which nuclear masses are evaluated without shell effects.

\begin{figure}
\includegraphics[width=9.6cm]{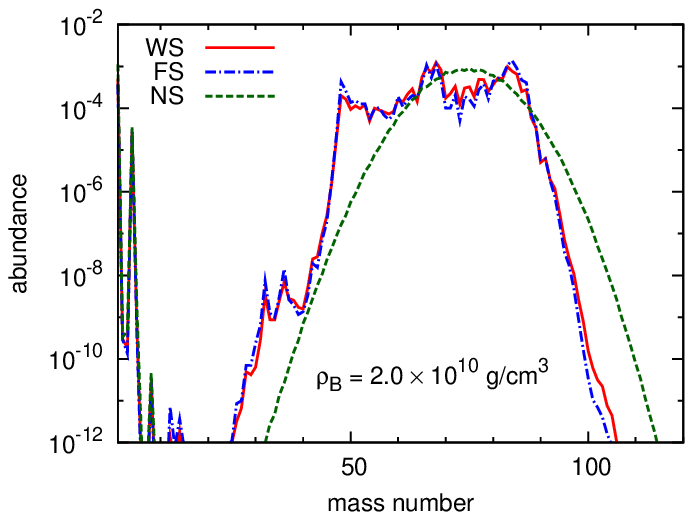}
\includegraphics[width=9.6cm]{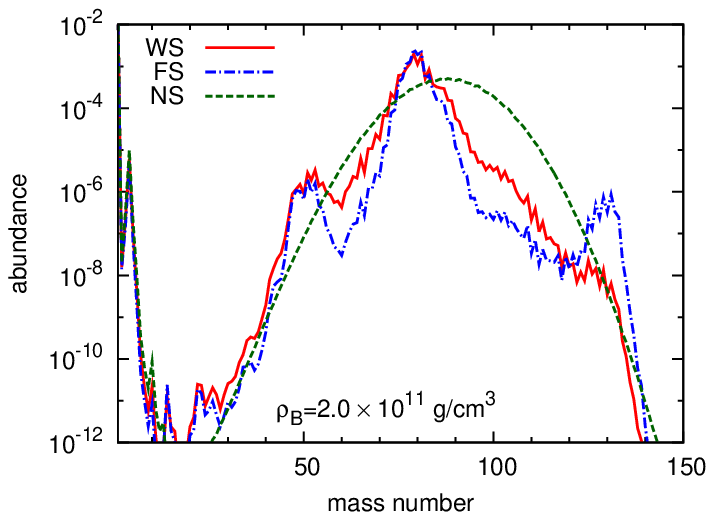}
\includegraphics[width=9.6cm]{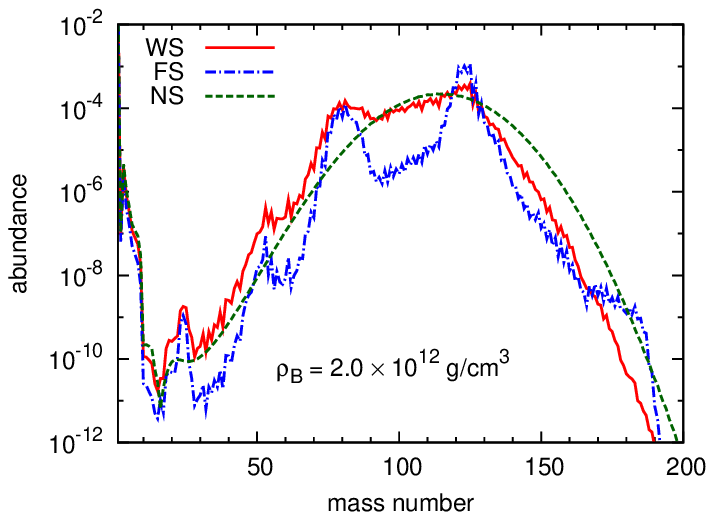}
\caption{
Abundances of elements as a function of the  mass number
for Models~WS  (red solid lines),~FS  (blue dashed-dotted lines) and NS  (green dashed lines) 
at ($\rho_B$~[g/cm$^3$], $T$~[MeV], $Y_e$)= (2.0 $\times 10^{10}$, 0.63, 0.41),
 (2.0 $\times 10^{11}$, 0.90, 0.36) and  (2.0 $\times 10^{12}$, 1.25, 0.29)
from top to bottom.}
\label{fig_mas1012}
\end{figure}

\begin{figure}
\includegraphics[width=9cm]{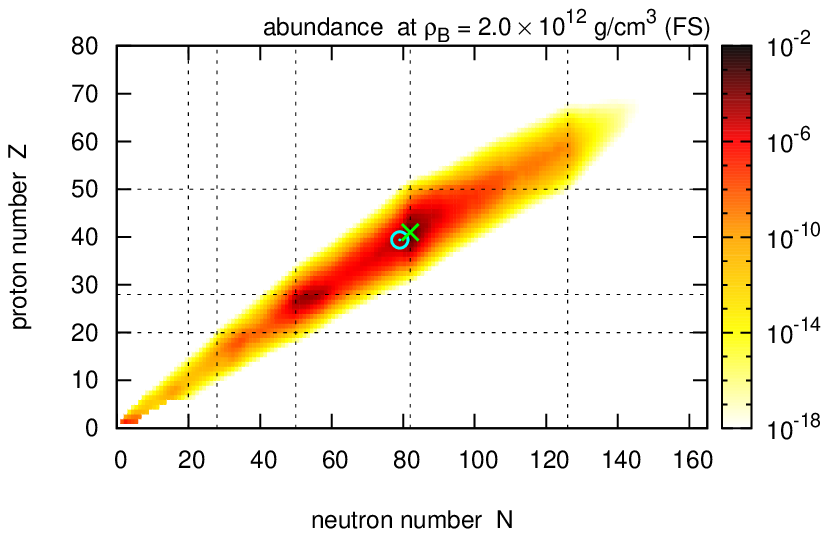}
\includegraphics[width=9cm]{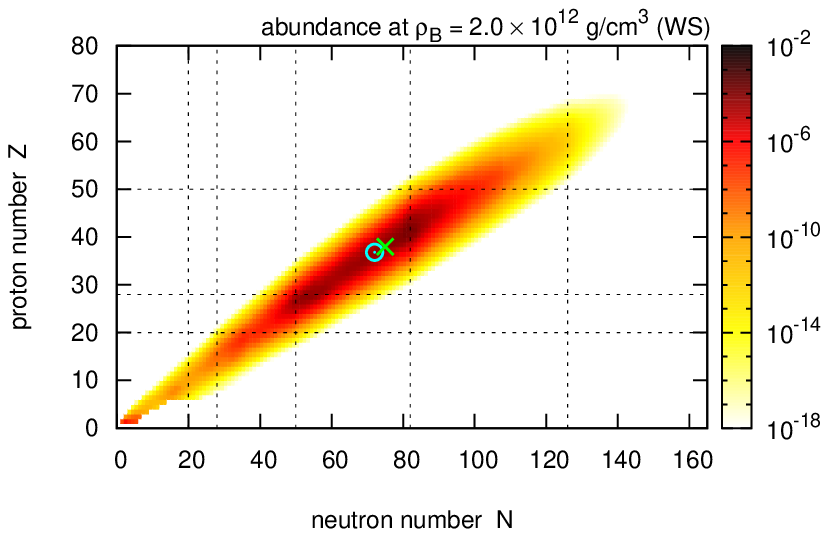}
\includegraphics[width=9cm]{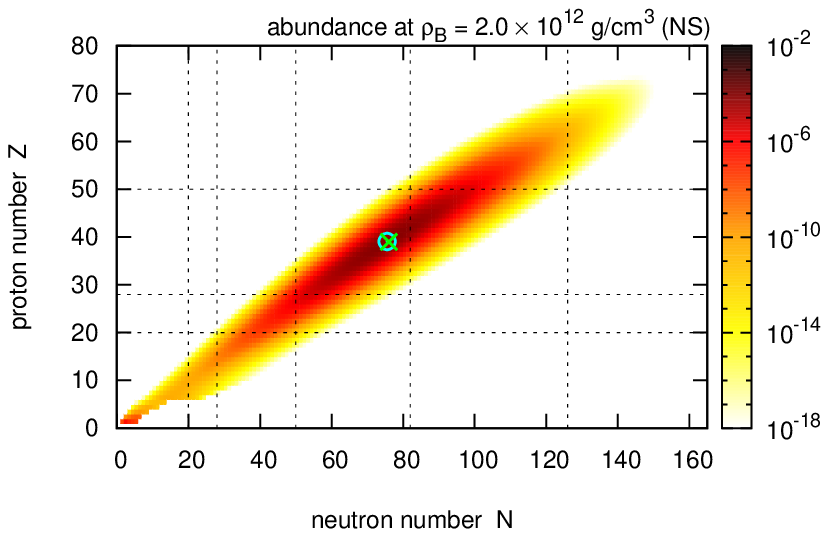}
\caption{
Nuclear abundances in the $(N, Z)$ plane at  $\rho_B=2.0 \times 10^{12}$~g/cm$^{3}$, $T=1.25$~MeV, and $Y_e = 0.29$  for Models  FS (top),  WS (middle) and NS (bottom)  from top to bottom.
The  cyan circled dot and  green cross indicate the  average values
and most probable nucleus, respectively.  
Dotted lines are neutron and proton  magic numbers.
}
\label{fig_dis12cde}
\end{figure}

We calculate nuclear compositions by the fragment definitions for the thermodynamical states at the center of the collapsing core, which are given in Figure~\ref{fig_tye}. 
 Figure~\ref{fig_mas1012} shows the abundances of elements as a function of  the mass number at ($\rho_B$~[g/cm$^3$], $T$~[MeV], $Y_e$)= (2.0 $\times 10^{10}$, 0.63, 0.41),
 (2.0 $\times 10^{11}$, 0.90, 0.36) and  (2.0 $\times 10^{12}$, 1.25, 0.29),
 which  are defined as $\displaystyle{\sum_{Z_i+N_i=A} n_i/n_B}$, 
 where $n_B$ is the baryon number density.
We can see from the comparison of  Models FS and WS  that
 the washout effect has no influence on the nuclear composition 
at $\rho_B= 2.0 \times 10^{10}$  g/cm$^{3}$,
whereas
the mass distributions in Model WS  are smoother compared to those in Model FS 
at $\rho_B= 2.0 \times 10^{11}$  and 10$^{12}$ g/cm$^{3}$.
We can also  find that this effect
reduces the sharpness of  the peaks at the neutron magic numbers $N =28, 50$ and $82$ ($A \sim 50, 80$ and 130) in the element
distributions; especially the abundances around the third peak of $N=82$
is reduced more  because of  the  smaller energy spacing, $\epsilon_{sh}$, for nuclei with the larger $A$
as discussed  in Furusawa et al. \cite{furusawa17a}.
Figure~\ref{fig_dis12cde} displays 
nuclear abundances, $n_i/n_B$, in the $(N, Z)$ plane for the three models  at $\rho_B= 2.0 \times 10^{12}$ ~g/cm$^{3}$.
 It is clear that nuclei are abundant in the vicinities of the neutron magic
numbers in Model FS.
On the other hand,  the nuclear distribution is much
smoother in Model NS  than that in Model FS.
The effect of shell washout in Model WS provides 
an intermediate distribution between those of Models FS and NS.

Figure~\ref{fig_massevo} shows the total mass fraction and average mass and proton numbers  for heavy nuclei  ($Z \geq 6$)   as functions of the central density.
They are 
 defined as $X_H=\sum_i A_i  n_i /n_B$, $\bar{A}=\sum_i A_i  n_i/\sum_i n_i$ and  $\bar{Z}=\sum_i Z_i  n_i/\sum_i n_i$. 
The mass and proton numbers of the most probable nucleus  in the ensemble of heavy nuclei, $A_{\rm mp}$ and $Z_{\rm mp}$, are also displayed.
 We find that the washout of shell effect reduces the mass fraction of heavy nuclei under the considered thermodynamical conditions,
whereas the reduction 
 is smaller than a few percent owing to the fact that  the closed-shell nuclei decrease
at the same time as the open-shell ones increase as shown in Figure~\ref{fig_dis12cde}.
The re-distribution of heavy nuclei, especially reduction of the peak nuclei, 
alters  chemical potentials of nucleons,
thereby affecting, a little, 
the balance between heavy nuclei and  the other baryons of nucleons and light nuclei.
Note that the total mass fraction of heavy nuclei is not necessarily reduced by the shell suppression \cite{furusawa17a}.
The average mass and proton numbers in Model WS do not  always settle
 down to values between those in Models FS and NS, since shell effects are 
sensitive to the nuclear species and the washout affects the average values non-linearly.
The mass and proton numbers of the most probable nucleus deviate from the average values in Models FS and WS due to shell effects,
 whereas they show close agreement in  Model NS.
Note that  
the neutrinos can barely escape from the core after the neutrino sphere is formed around $\rho_B \sim 2  \times 10^{12}$ ~g/cm$^{3}$.
Therefore, the lepton fraction of the core is little reduced by  weak interactions above this density
  \cite{sullivan16}.
We, hence, focus only  on the densities lower than  $\rho_B \sim 2 \times 10^{12}$ ~g/cm$^{3}$.

\section{Weak interaction rates \label{sec:weak}}
The electron capture rate of each nucleus is estimated by the weak rate tables of
 FFN \cite{fuller82}, ODA \cite{oda94}, LMP \cite{,langanke00} and  LMSH \cite{,langanke03}.
For the nuclei where no data are available,
 we utilize the following approximation formula  as a  function of the $Q$ value \cite{fuller85,langanke03}:
\begin{eqnarray}
 \lambda_i =\frac{({\rm ln}2)B}{K} \left( \frac{T}{m_e c^2} \right)^{5} \left[ F_4(\eta_i) - 2 \chi_i F_3(\eta_i) + \chi_i^2 F_2(\eta_i)  \right] , \label{eq:appro} 
\end{eqnarray}
where  $K=6146$ sec, $\chi_i=(Q_i - \Delta E)/T$, $\eta_i=(\mu_e +  Q_i -\Delta E)/T$ with the electron chemical potential $\mu_e$ and $F_k$ is the relativistic Fermi integral of order $k$.
The parameters of a typical matrix element  ($B=4.6$) and  a transition energy from an excited state in the parent nucleus to a daughter state ($\Delta E=2.5$ MeV) are fitted to shell-model calculations for
the pf-shell nuclei of LMP data by Langanke et al. \cite{langanke03}.
The $Q$ value of each nucleus, 
 $Q_i$,  is calculated by the mass formulae for heavy nuclei, which is introduced in the previous subsection,   
as $Q_i=M_i(Z_i,N_i)-M_i(Z_i-1,N_i+1) $ with in-medium effects  at finite density and temperature.
We adopt reaction rates in the predetermined order as
 LMP $>$ LMSH  $>$ODA $>$FFN $>$ approximation formula, which means that rates from sources with higher orders are utilized
for nuclei whose rates from multiple sources exist. 
Figure~\ref{fig_nuclide} shows the sources in nuclear chart,
 which are applied to 
each nuclei. 
The light nuclei ($Z \leq 5$) are ignored in this work, since they are not abundant
 under the considered thermodynamical conditions.

The neutrino-nucleus scattering rate is calculated  for electron-type neutrinos with the  average energy, $E_{\nu_e}$,
whose values are taken from the result of  the reference simulation and  shown in Figure~\ref{fig_tye}.
The cross section of an  individual nucleus is evaluated as
\begin{eqnarray}
 \sigma_i (E_{\nu_e}) =  \frac{G_W^2}{8 \pi ( \hbar c)^4} E_{\nu_e}^2 A_i^2 \left\{ 1- \frac{2 Z_i}{A_i}(1-{\rm sin}^2 \theta_W)  \right\}^2 \frac{2 y_i  + {\rm exp}(2 y_i) -1}{y_i^2} ,
 \label{eq:sca}
\end{eqnarray}
where $y_i= 1.92 \times 10^{-5} A_i^{2/3} E_{\nu_e}^2$,   $G_W$ and $\theta_W$ are  the weak coupling constant and  Weinberg angle, respectively,  
and the isoenergetic  zero-momentum transfer and non-degenerated  nucleus are assumed   \cite{bruenn85}.

\subsection{Dependence of  Weak Interaction Rates on Shell Effect}
\begin{figure}
\includegraphics[width=12cm]{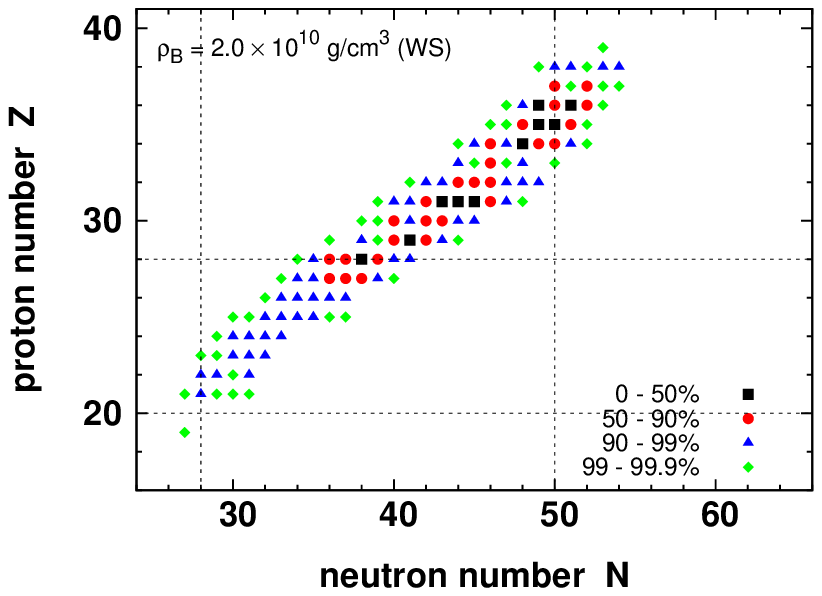}
\includegraphics[width=12cm]{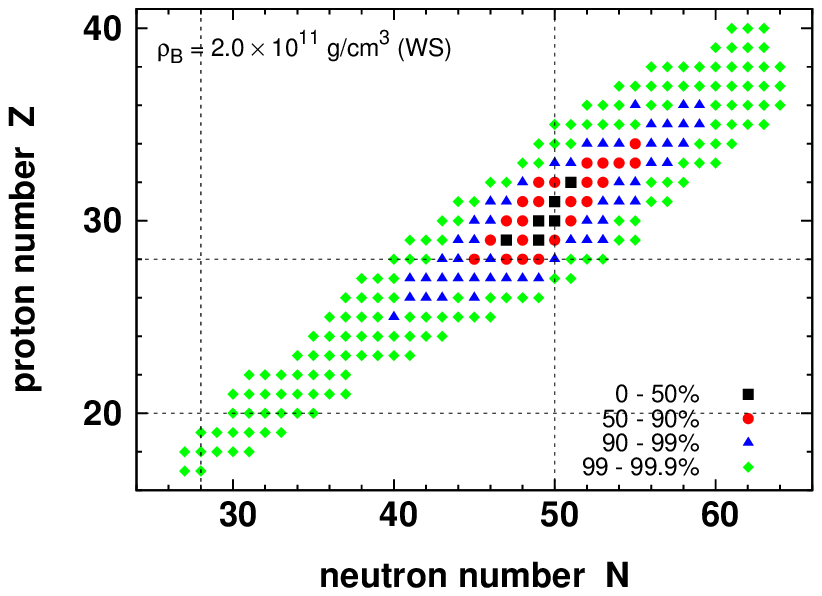}
\caption{The nuclear species with the largest electron capture contribution, $n_i \lambda_i/n_B$, 
which account for the top 50$\%$ (black squares), 50-90$\%$ (red circles), 90-99$\%$ (blue triangles)   and 99-99.9$\%$ (green diamonds) 
 of  the total electron capture rate per baryon $\lambda^{ec}$
 for  Model WS 
at  $\rho_B= 2.0 \times 10^{10}$~g/cm$^{3}$ (top panel) and
$2.0 \times 10^{11}$~g/cm$^{3}$ (bottom panel). 
Dotted lines are neutron and proton  magic numbers.
}
\label{fig_ele1011}
\end{figure}

\begin{figure}
\includegraphics[width=9.5cm]{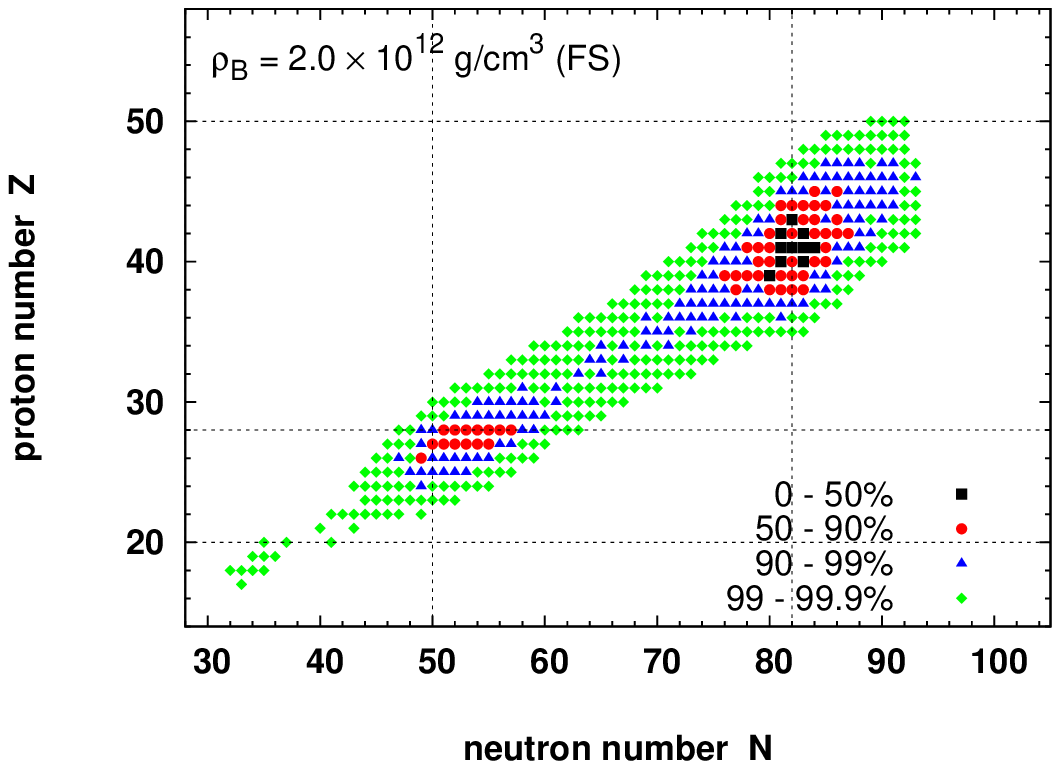}
\includegraphics[width=9.5cm]{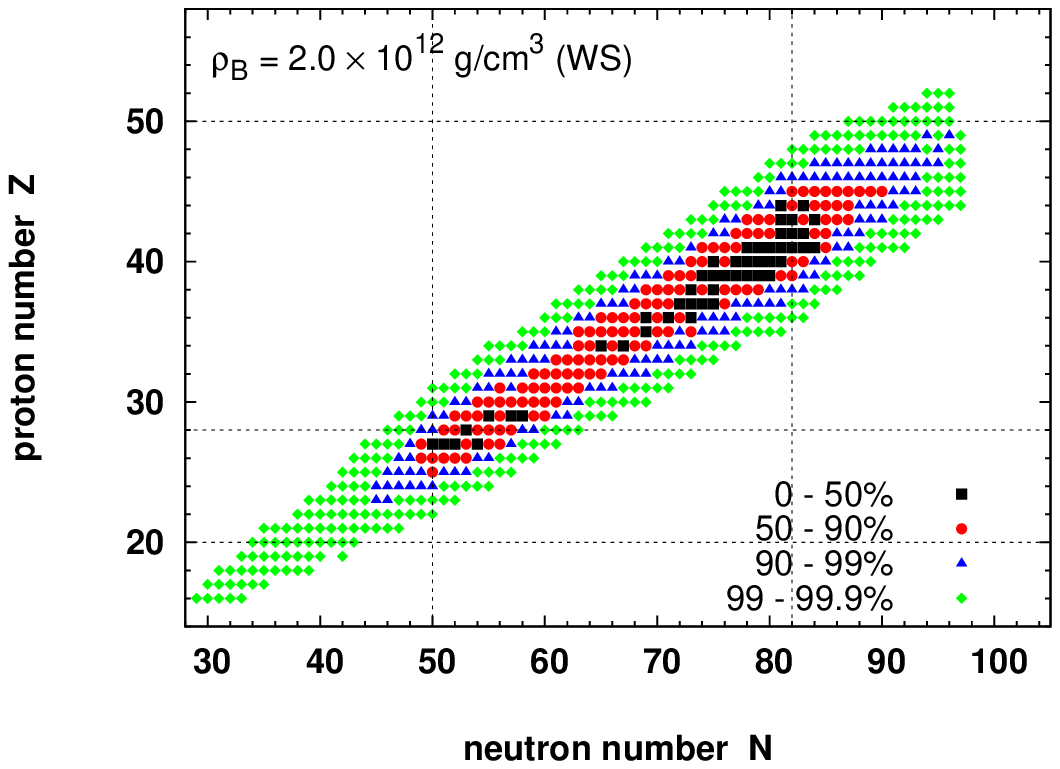}
\includegraphics[width=9.5cm]{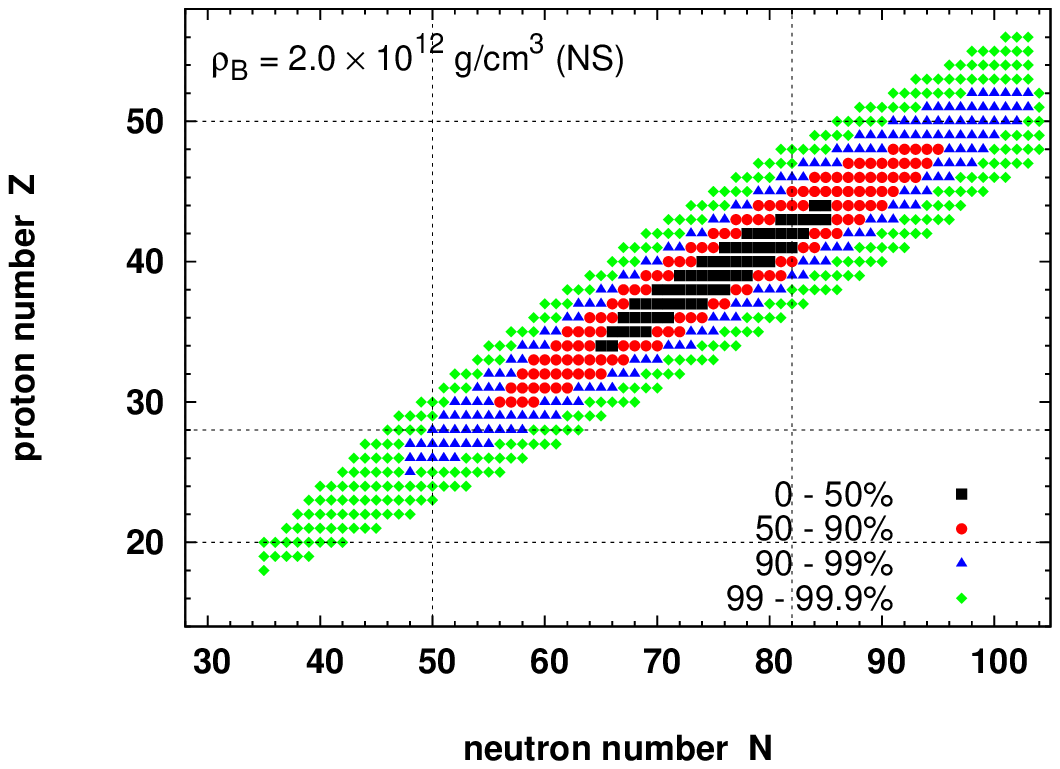}
\caption{
The nuclear species   with the largest electron capture contribution, $n_i \lambda_i/n_B$, 
which  account for the  top 50$\%$ (black squares), 50-90$\%$ (red circles), 90-99$\%$ (blue triangles) and 99-99.9$\%$ (green diamonds)
   of  the total electron capture  rate per baryon $\lambda^{ec}$
for Models  FS (top),  WS (middle) and NS (bottom) 
at  $\rho_B=2.0 \times 10^{12}$~g/cm$^{3}$.
Dotted lines are neutron and proton  magic numbers.
}
\label{fig_elecde12}
\end{figure}

We discuss the contribution of each nucleus to the total electron capture rate per baryon defined as $\lambda^{ec}=\sum_i  n_i \lambda_i/n_B$.
Figure~\ref{fig_ele1011} shows 
the nuclei with the largest contributions, $n_i \lambda_i/n_B$, which make the top 50$\%$, 90$\%$, 99$\%$ and 99.9$\%$ of $\lambda^{ec}$ for  Model WS 
 at $\rho_B= 2.0 \times 10^{10}$ and  $2.0 \times 10^{11}$~g/cm$^{3}$. 
At the beginning of collapse ($\rho_B= 2.0 \times 10^{10}$~g/cm$^{3}$),  the nuclei with $36 \leq N \leq 52$  account for the top 90$\%$ of the total electron capture.
On the other hand, 
the nuclei around the second peak ($45\leq N \leq 55$) make up
$90\%$  at  $\rho_B= 2.0 \times 10^{11}$~g/cm$^{3}$.
Figure~\ref{fig_elecde12} compares
the nuclei  with large contributions 
 at $\rho_B= 2.0 \times 10^{12}$~g/cm$^{3}$ for all models. 
The dominant nuclei  correspond  more or less to the nuclei with large abundances in Figure~\ref{fig_dis12cde}.
For Model FS,  
two islands around  $N=50$ and $82$  are clearly visible in the distribution due to shell effects.
This feature has also been observed in the previous work \cite{sullivan16}, in which the shell washout is neglected.
On the other hand, the distributions of dominant nuclei for Models WS and NS are more broad
and the non-magic nuclei such as $^{105}$Kr ($N=69$)  also contribute to the total rate.
 We also find that the numbers  of the  nuclei with large contributions  increase
 as the shell effects become weaker; 
these 
 are 10, 48 and 59 for the top 50$\%$ and  62, 175 and  188  for  the top 90$\%$  in Models FS, WS and  NS, respectively.  
 
\begin{figure}
\includegraphics[width=13cm]{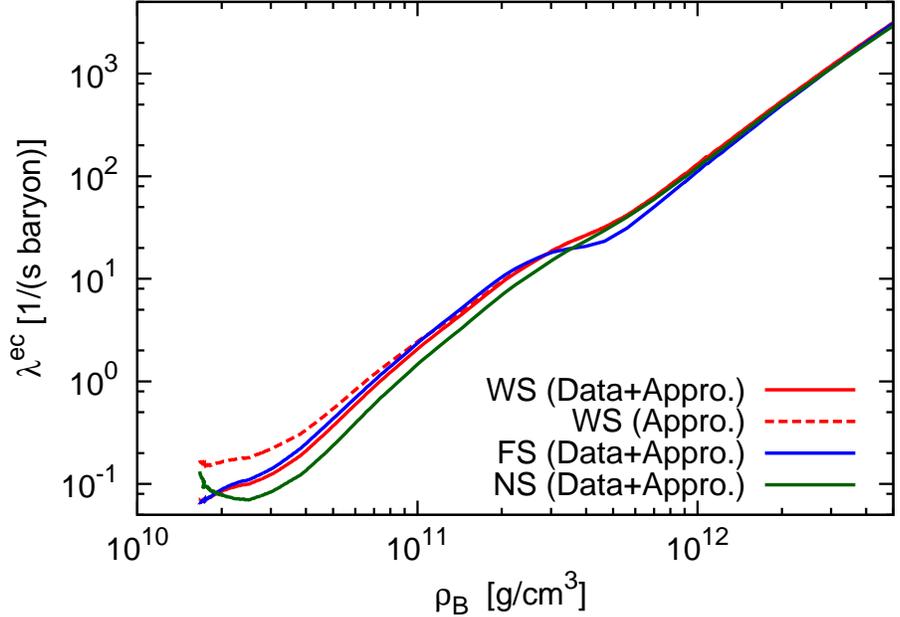}
\caption{
Electron capture rate per baryon  of  heavy nuclei, $\lambda^{ec}=\sum_i  n_i \lambda_i/n_B$,  as a function of central density  for  Models~WS (red solid lines),
 FS  (blue solid lines) and NS  (green solid lines). 
The red dashed line shows the result for Model WS in which
the  approximation formula   is adopted   for all nuclei  and weak rate data are not utilized. }
\label{fig_elecde}
\end{figure}

 Figure~\ref{fig_elecde} displays 
the electron capture rate of  heavy nuclei  per baryon $\lambda^{ec}$  as a function of central density, 
 which corresponds to the time derivative of electron fraction  in collapsing cores by
 the reactions, ${\rm d} Y_e/ {\rm d} t$. Note that neutrino blocking is not considered here. 
 The weak rate data (FFN, ODA, LMP and LMSH) are used whenever available
  in Models FS, WS  and NS  (Data+Appro.), whereas the approximation formula of Eq.~(\ref{eq:appro}) is applied to
 all nuclei in the model WS (Appro.).
We can 
see from the comparison of  models WS with and without the data,
that  the weak rate data are influential at  densities below $\rho_B \sim 10^{11}$~g/cm$^3$;  
this is because the nuclei  that are not included in these data become abundant at high densities.

\begin{figure}
\includegraphics[width=10cm]{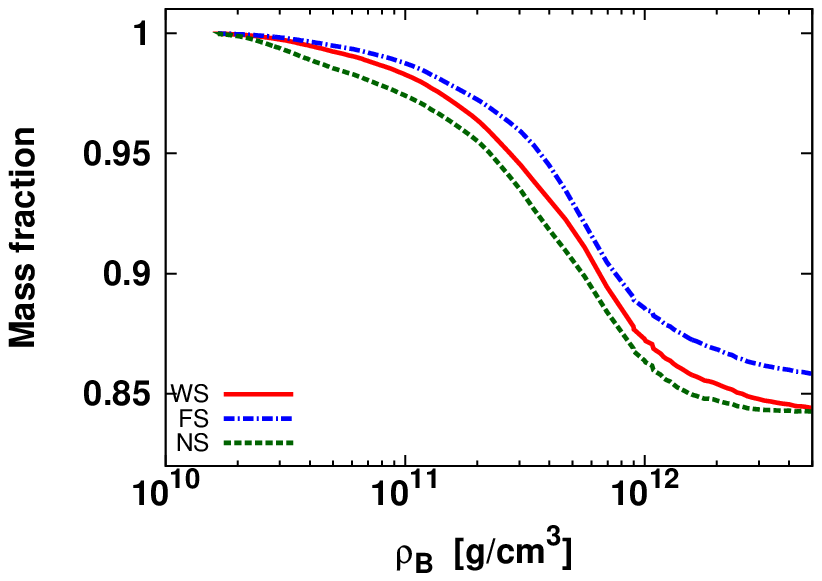}
\includegraphics[width=10cm]{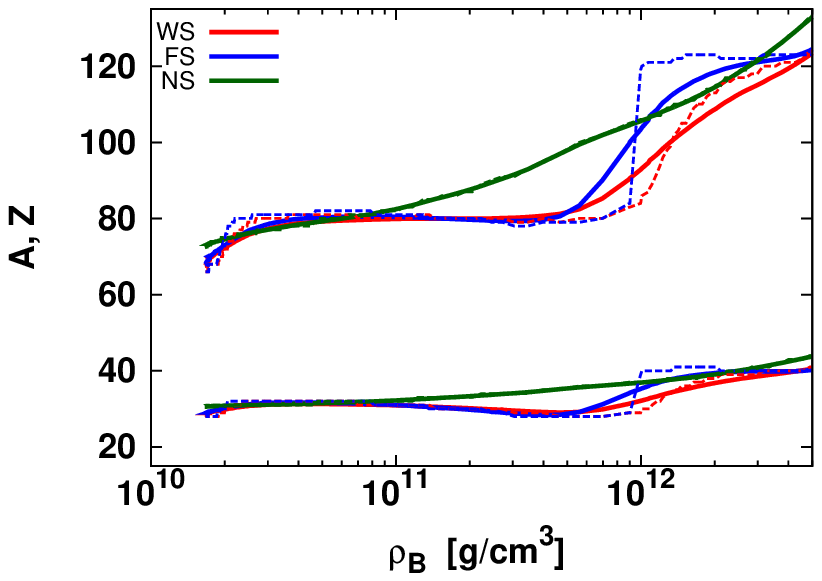}
\caption{
Mass fraction of heavy nuclei with $Z \geq 6 $ as a function of central density for Models~WS  (red solid lines),~FS  (blue dashed-dotted lines) and NS  (green dashed lines)  in top panel. 
The bottom panel displays the  average mass and proton numbers, $\overline{A}$ and $\overline{Z}$, (solid lines)  and those of the most probable nuclei,  $A_{\rm mp}$ and $Z_{\rm mp}$,  (dashed lines) 
for Models~WS  (red),~FS  (blue) and NS  (green).}
\label{fig_massevo}
\end{figure}

\begin{figure}
\includegraphics[width=12cm]{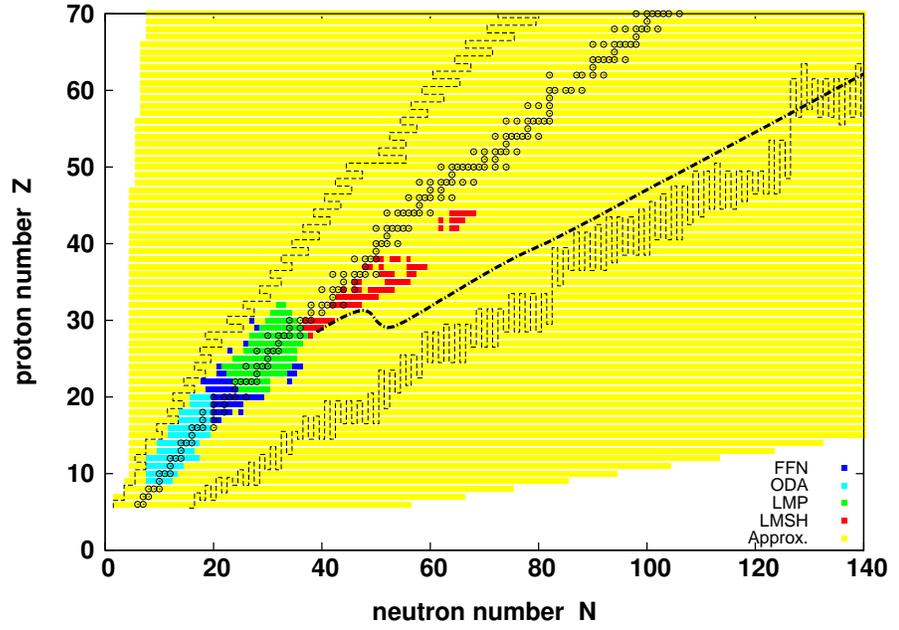}
\caption{ The nuclear species, for which the data of FFN (blue squares),
ODA (cyan squares), LMP (green squares), or LMSH (red squares),  or approximation formula (yellow squares)  for electron capture rate is adopted. 
Black circles represent stable nuclei and dashed black lines  display 
 the neutron and proton drip lines (dashed black lines), which are estimated by the KTUY mass formula \cite{koura05}. 
 A dashed-dotted line indicates the trajectory of  
average neutron and proton numbers of heavy nuclei 
 in  the center of  collapsing cores, which is calculated by Model WS
and shown in Figure~\ref{fig_massevo}
.}
\label{fig_nuclide}
\end{figure}

\begin{figure}
\includegraphics[width=8cm]{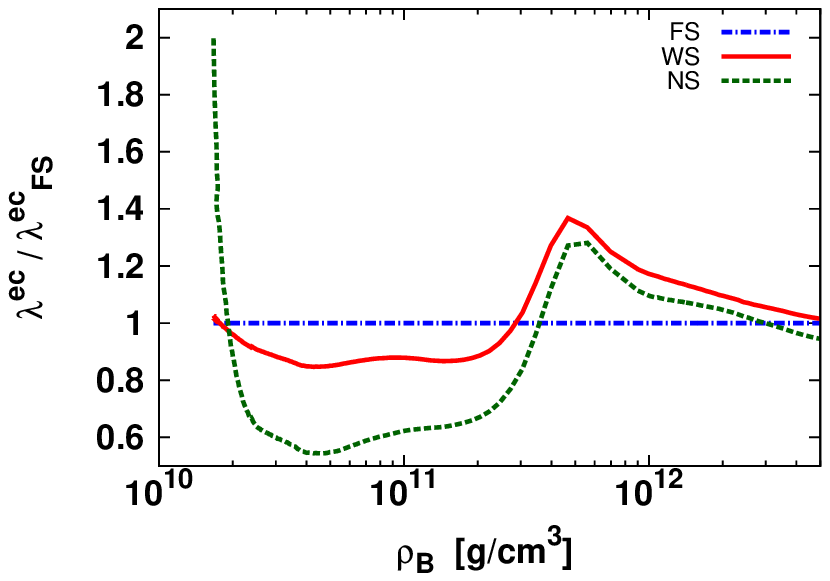}
\includegraphics[width=8cm]{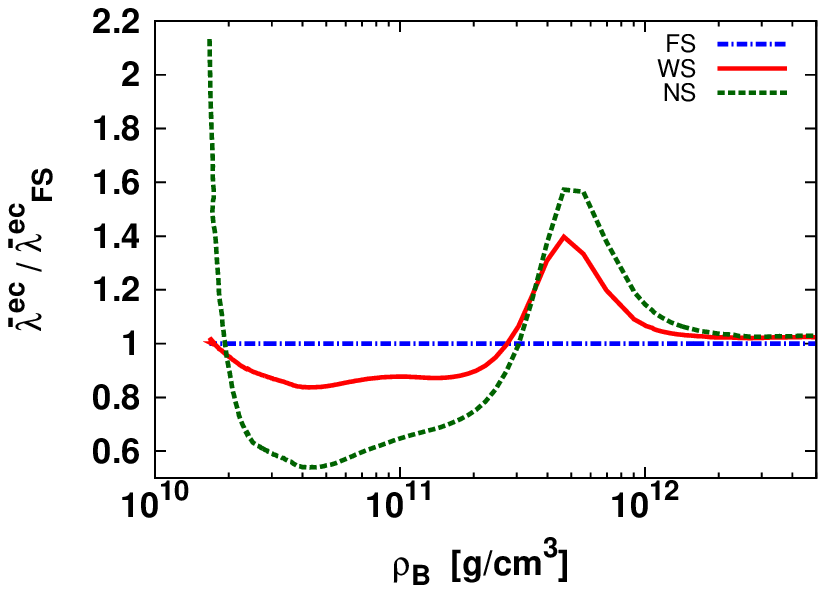}
\caption{Ratio of electron capture rates per baryon $\lambda^{ec}$ (left panel) and those per nuclei, $\bar{\lambda}^{ec}= \sum_i  n_i \lambda_i/(\sum_i  n_i)$, (right panel)
for Models~WS  (red solid lines) and NS  (green dashed lines) 
 to those for  Model~FS  (blue dashed-dotted lines)
 as a function of central density. 
}
\label{fig_ratioelecde}
\end{figure}

Figure~\ref{fig_ratioelecde} shows the ratio of electron capture rates for Models WS and NS  to those for Model FS.
The left panel compares  the rates per baryon $\lambda^{ec}$,
while the right panel displays those per nuclei $\bar{\lambda}^{ec}=\sum_i  n_i \lambda_i/(\sum_i  n_i)$.
They can be converted to each other as  $ \lambda^{ec}= \bar{\lambda}^{ec} X_H/ \bar{A}$
 by using 
 $X_H$ and 
 $\bar{A}$, which are shown in Figure~\ref{fig_massevo}.
We find that the shell smearing cuts down on $\lambda^{ec}$ and $\bar{\lambda}^{ec}$ by $ \sim 20\%$ at low densities and raises them by $\sim 40\%$ at high densities.
At around  $\rho_B = 4 \times 10^{10}$~g/cm$^3$, Model FS displays the  largest rates, since   the mass fractions of the nuclei around the first peak with $N=28$  are the largest among the three models. 
These nuclei  at the first peak of $A \sim 50$ ($N \sim 28$)  have higher
reaction rates than those of  the nuclei with larger mass numbers.
The reduction of nuclear abundance at  the first peak  results in  lower electron capture rates   in Model WS than those in Model FS. 
At  densities greater than $\rho_B \sim   3 \times 10^{11}$~g/cm$^3$, Model WS  yields higher rates than Model FS,
since the magic nuclei are less  and  non-magic ones are more abundant  as shown in Figures~\ref{fig_mas1012},  \ref{fig_dis12cde} and~\ref{fig_elecde12}.
This feature is essentially similar to the result observed in  Raduta et al.  \cite{raduta16} in that shell quenching reduces the closed-shell nuclei and increases the  averaged electron capture rate. 
Model  WS  has  larger $\lambda^{ec}$ than Model NS because of larger nuclear abundances $\sum_i n_i$ (larger $X_H$ and smaller $\bar{A}$ as shown in Figure~\ref{fig_massevo}).
On the other hand, the rates per nuclei, $\bar{\lambda}^{ec}$, 
 in Model WS
 settle down to values between those in Models NS and FS.
At  densities greater than $\rho_B \sim 2 \times 10^{12}$~g/cm$^3$, all models lead to the same value of $\bar{\lambda}^{ec}$,
since $\mu_e$ becomes much larger than the  $Q$ values of nuclei and, as a result, their differences among nuclei become negligible.

\begin{figure}
\includegraphics[width=8.1cm]{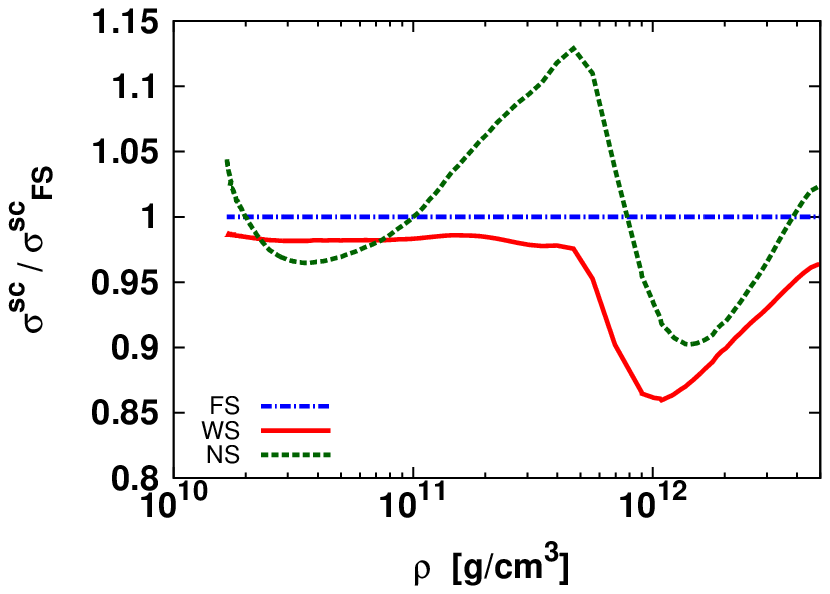}
\includegraphics[width=8.1cm]{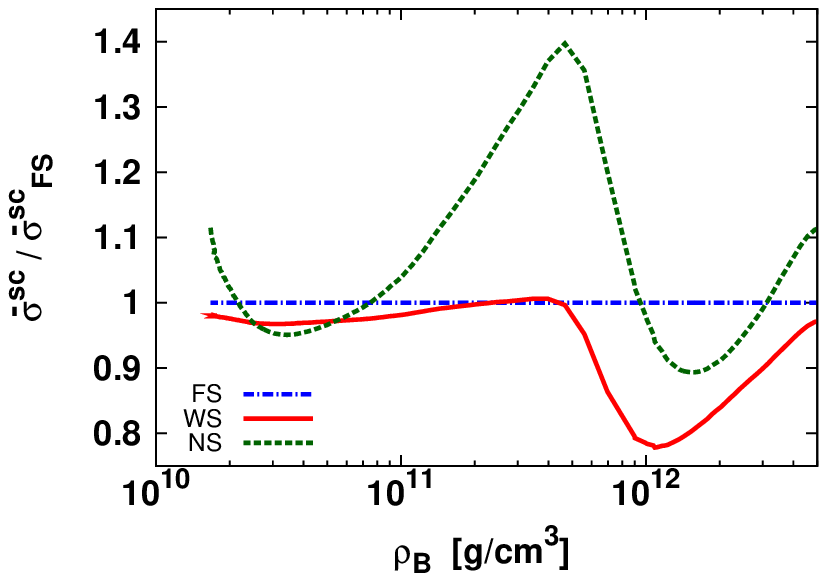}
\caption{Ratio of cross sections of neutrino-nucleus scattering 
per baryon  $\sigma^{sc}=\sum_i n_i \sigma_i/n_B$ (left panel) and those per nuclei $\bar{\sigma}^{sc}= \sum_i  n_i \sigma_i/(\sum_i  n_i)$ (right panel)
for Models~WS  (red solid lines) and NS  (green dashed lines) 
 to those for  Model~FS  (blue dashed-dotted lines)
 as a function of central density. }
\label{fig_ratioscatcde}
\end{figure}

Figure~\ref{fig_ratioscatcde} displays the ratio of the cross sections of neutrino-nucleus scattering in Models WS and NS to those in Model FS.   Left  and right panels compare  the cross sections per baryon $\sigma^{sc}=\sum_i  n_i \sigma_i/n_B$
and those per nuclei $\bar{\sigma}^{sc}=\sum_i  n_i \sigma_i/\sum_i  n_i$, respectively.
The proportional relation in Eq.~(\ref{eq:sca}), $\sigma_i \appropto A_i^2$, 
 gives  the 
approximate relations of those rates as: $\sigma^{sc} \appropto X_H  \overline{A^2}/\overline{A} \sim X_H  \overline{A} $ and $\bar{\sigma}^{sc}    \appropto \overline{A^2} \sim \overline{A}^2$, where  $\overline{A^2}=\sum_i A_i^2  n_i/\sum_i n_i$.
The difference in $X_H$ is not very great as shown in Figure~\ref{fig_massevo}
and, hence, models with larger  $\bar{A}$  lead to larger  $\sigma^{sc}$
and $\bar{\sigma}^{sc}$.
For instance, Model NS gives the greatest cross section at  $\rho_B \sim 4 \times 10^{11}$~g/cm$^3$.
We find that the washout of the shell effect reduces $\sigma^{sc}$ by $\sim 15\%$ around 
 $\rho_B \sim 10^{12}$~g/cm$^3$, since the nuclear abundances with large mass numbers decrease and $\overline{A}$ becomes small.
The cross sections per nuclei, $\bar{\sigma}^{sc}$, are reduced more than $\sigma^{sc}$ and down about 20$\%$, since they are proportional  to $A^2$.  

\subsection{Approximation Errors of  Single Nucleus Description in Weak Rates}
As explained in the Introduction, the single-nucleus EOS has been utilized in most supernova simulations, 
where the weak rates of the representative nucleus are substituted for the exact rates
obtained by folding individual rates with NSE abundances.
Even in the simulation with multi-nucleus EOS,  the average values  such as $\overline{A}$ and $\overline{Z}$ are utilized  to estimate the neutrino-nucleus scattering
 just in the same way as in the single nucleus description.
The same applies to the reference simulation of core collapse.
In this subsection, we discuss the errors in the weak rates of the single nucleus descriptions. 

\begin{figure}
\includegraphics[width=11cm]{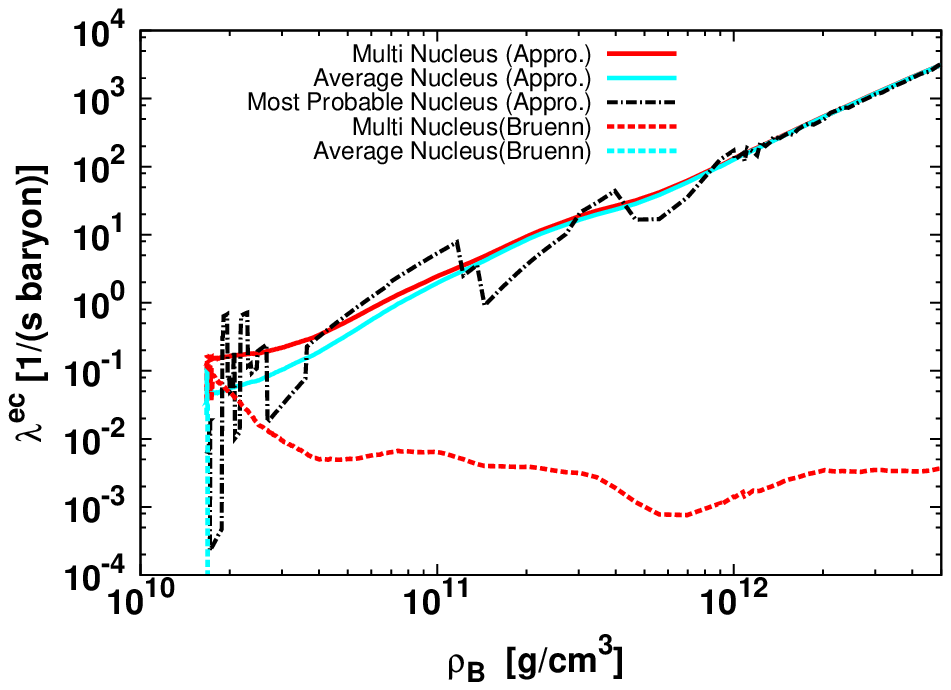}
\caption{
Electron capture rate per baryon of  heavy nuclei   as a function of central density for  Model WS  based on the multi-nucleus description with the parametrized rate provided by Langanke et al.  \cite{langanke03}    (red solid lines)  and the Bruenn rate \cite{bruenn85} 
(red dashed lines). 
Those  of the average nucleus 
are shown
for the former rate (cyan solid lines)  and the latter  (cyan dashed lines).
The black dashed-dotted line shows that of  the most probable nucleus for the former rate.
}
\label{fig_elesm}
\includegraphics[width=8.1cm]{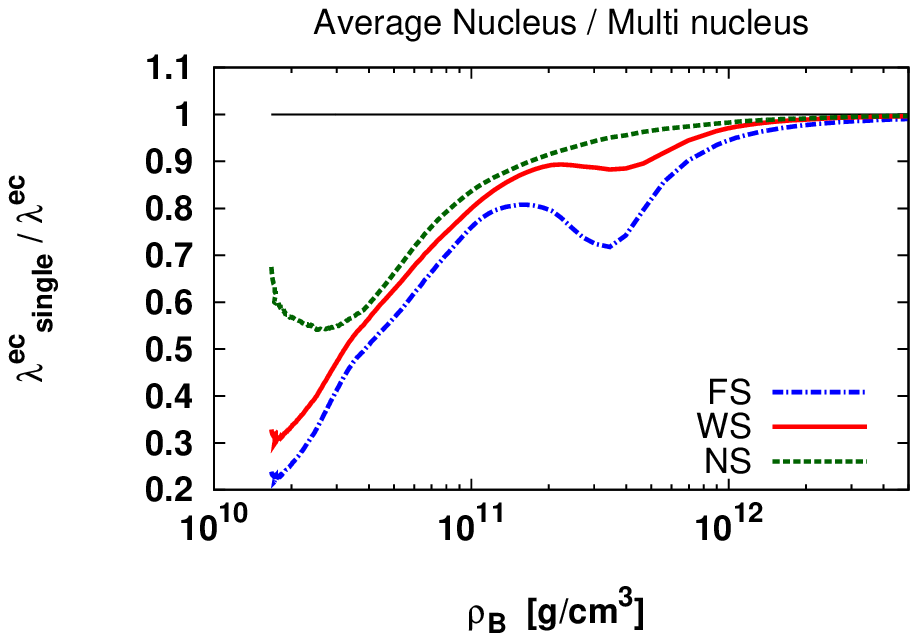}
\includegraphics[width=8.1cm]{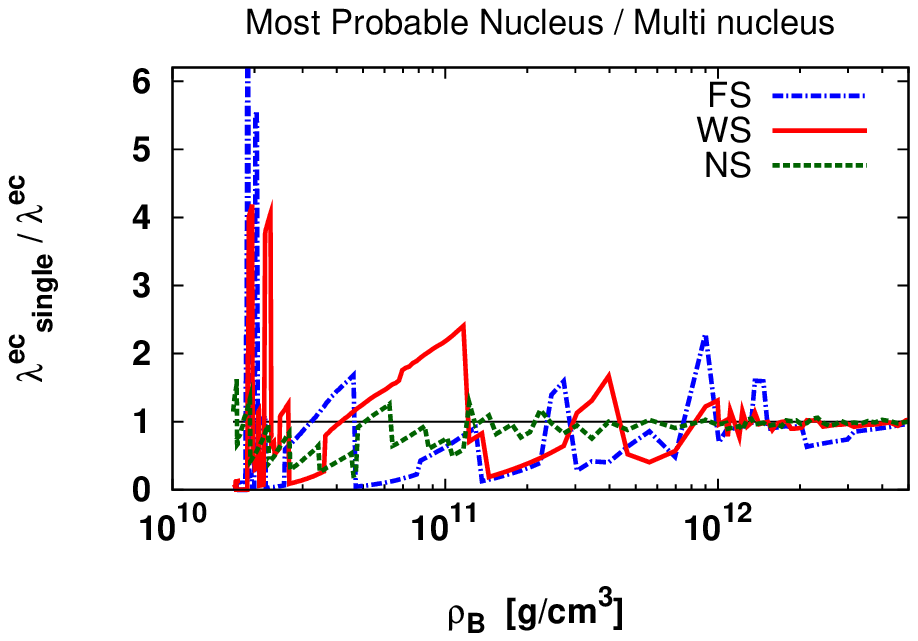}
\caption{
 Ratio of electron capture rates  in the single-nucleus descriptions  to those of the multi-nucleus descriptions
 for  Models~WS (red solid lines), FS  (blue solid lines) and NS  (green solid lines).
Left and right panels display those of the average nucleus 
 and the most probable one. }
\label{fig_elesmcmp}
\end{figure}

We estimate the electron capture rates per baryon, $\lambda^{ec}_{\rm single}$, 
for the average nucleus and  the most probable one in the single-nucleus descriptions,
in which the representative nucleus  is assumed to account for the total mass fraction of heavy nuclei $X_H$ alone.
 The former is expressed as  $\lambda^{ec}_{\rm single} (\overline{Q},\overline{A})=  X_H/\overline{A} \  \lambda(\overline{Q})$ with Eq.~(\ref{eq:appro})
on the assumption that the average nucleus has $\overline{A}$, $\overline{Z}$ and  the average $Q$ value, $\overline{Q}$.
The latter is defined as $\lambda^{ec}_{\rm single} (Q_{\rm mp}, A_{\rm mp})= X_H/A_{\rm mp} \lambda (Q_{\rm mp})$ 
using the individual rate, $\lambda (Q_{\rm mp})$, and mass number, $A_{\rm mp}$, of the most probable nucleus. 
In the multi-nucleus description,  Eq.~(\ref{eq:appro}) is applied to all nuclei as $\lambda^{ec}=\sum_i n_i \lambda(Q_i)/n_B$. 
 Note that weak rate tables are not used here for simplicity.
Figure~\ref{fig_elesm} shows  these rates for Model WS
and 
Figure~\ref{fig_elesmcmp}  displays
the ratio of  $\lambda^{ec}_{\rm single}$ to  $\lambda^{ec}$ for all models. 
It is clear to see that  the electron capture rates based on the average values are smaller
due to the neglect of the nuclei other than those at abundance peaks.
The nuclei  with smaller mass numbers and/or  larger charge fractions $Z_i/A_i$, which are not included  in the single nucleus description, have
 reaction rates higher than those of the nuclei at the abundance peak.
This artificial error
 in Model FS is the largest  among the three  models, 
since the average values are the closest in value to those of  the magic nuclei as shown in Figure~\ref{fig_dis12cde}, whose rates are lower than  those of non-magic ones.
We comment that the electron capture rates of the most probable nuclei
are discrete, since $A_{\rm mp}$ and $Z_{\rm mp}$
 adopt integer values and also $Q_{\rm mp}$ is not continuous either.
They are underestimated more often than not, but basically do not follow the trend of average values 
especially in  Models FS and WS because of  shell effects.

Figure~\ref{fig_elesm} also shows
the electron capture rates based on the old  formula provided by Bruenn  \cite{bruenn85}.
In this estimation,  the reaction rate is set to be zero for  the nuclei with  $N \geq 40$  as already noted.
We find that the formula is  out of the question.
The rate of average nucleus drops to zero  just after the core-collapse starts,
since the average neutron number exceeds 40.
Even in the multi-nucleus description with the old formula, the rate decreases as 
the nuclei with $N<40$ diminish
at high densities.

\begin{figure}
\includegraphics[width=8.1cm]{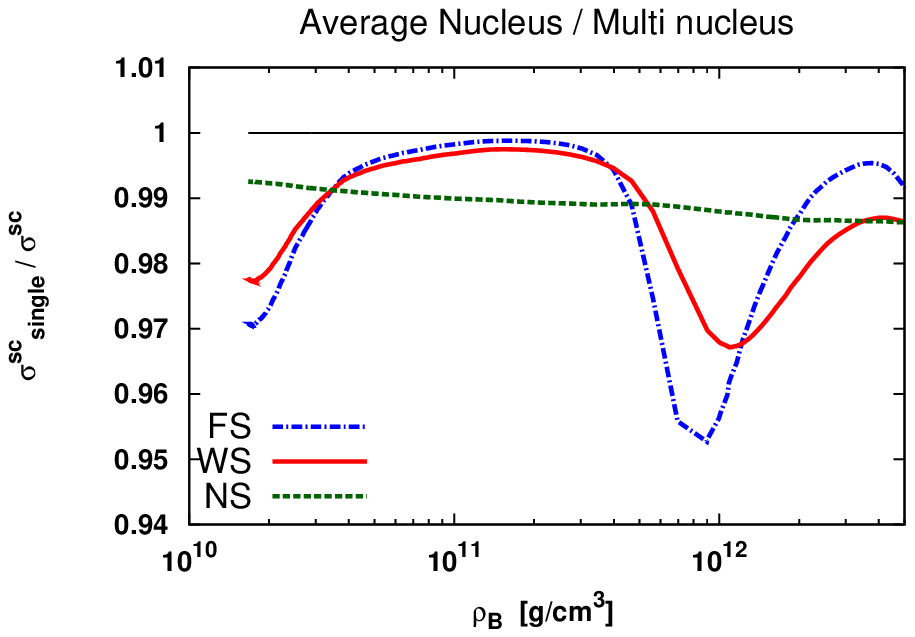}
\includegraphics[width=8.1cm]{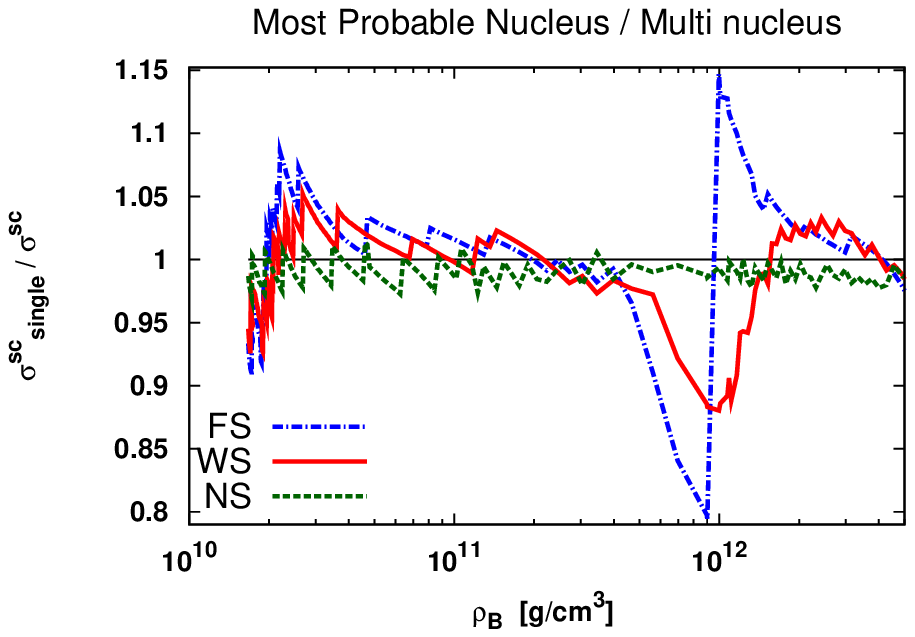}
\caption{
Ratio of cross sections of neutrino-nucleus scattering 
in the single-nucleus descriptions  to those of the multi-nucleus descriptions 
 for  Models~WS (red solid lines), FS  (blue solid lines) and NS  (green solid lines).
Left and right panels display those of the average nucleus 
 and the most probable one. }
\label{fig_scatsm}
\end{figure}

\begin{figure}
\includegraphics[width=12cm]{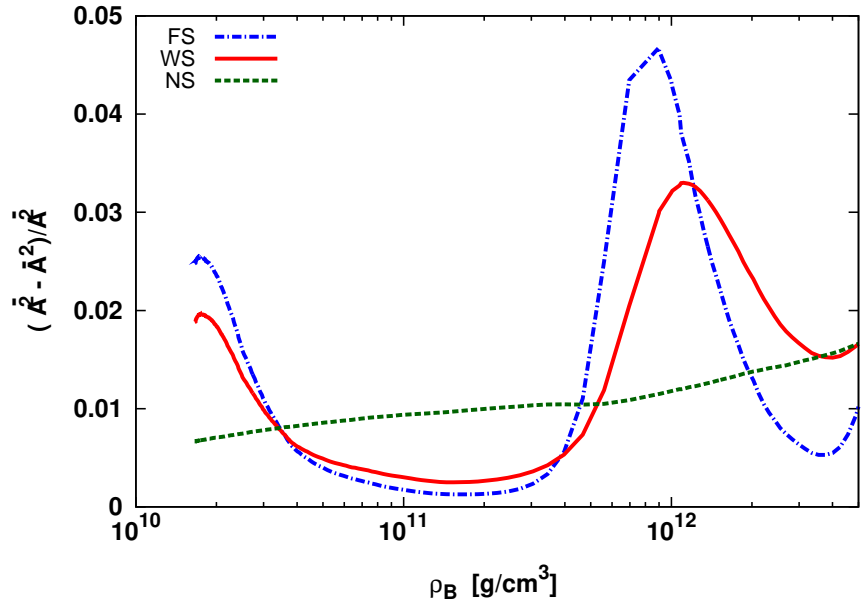}
\caption{Dispersion of mass number normalized by the average mass number squared, $(\overline{A^2}- \overline{A}^2)/\overline{A^2}$, as  functions of central density
for Models FS (blue dashed-dotted line), WS (red solid line) and NS (green dashed line).
}
\label{fig_disper}
\end{figure}

Finally we compare the cross section per baryon for neutrino coherent scattering on heavy nuclei  among the average nucleus
 ($\sigma^{sc}_{\rm single} (\bar{A}, \bar{Z})= \sigma(\bar{A}, \bar{Z}) X_H /\bar{A}$),
 the most probable one ($\sigma^{sc}_{\rm single} (A_{\rm mp}, Z_{\rm mp})= \sigma(A_{\rm mp}, Z_{\rm mp}) X_H /A_{\rm mp}$) and
the  multi-nucleus description ($\sigma^{sc}=\sum_i \{ n_i  \sigma(A_i, Z_i)  \}/n_B$) 
in  Figure~\ref{fig_scatsm}.
Unlike the  electron capture, the scattering is not sensitive to the feature of the individual nucleus and  its rate is roughly  proportional to $A^2$.
Hence, the approximation errors are not significantly large compared with electron captures.
We find that the errors in the single nucleus description based on the average nucleus depend on the dispersion of mass numbers, $\overline{A^2} -\bar{A}^2$,
as shown in Figure~\ref{fig_disper}.
In Models FS and WS, the dispersions  are small,  around  $\rho_B \sim 10^{11}$~g/cm$^3$,  
since closed-shell nuclei with $N=50$ dominate in the nuclear abundance. 
They grow  steeply around  $\rho_B \sim 10^{12}$~g/cm$^3$,
since the nuclei at the third peak $N=82$ 
appear
 and the average mass number rises as shown in Fig.\ref{fig_massevo}.
In Model WS, the shell washout reduces the dispersion and, as a result,  the approximation error is reduced.
The errors are about 1$\%$, 3$\%$ and  5$\%$ at most for Models NS, WS and FS, respectively. 
 The approximation errors  for the most probable nuclei  are larger than those for  the average  ones,
 although the deviations are  smaller than   those seen in the case of electron captures.

\section{Summary and Discussion \label{sec:sum}}
We have calculated 
the weak interaction rates of heavy nuclei in the collapsing core of the massive star
to investigate their sensitivities  to uncertainties in nuclear composition.
The abundance of various nuclei 
 is evaluated
 by the three different  EOS models.
One is our new EOS model including
  the washout of shell effect, which has been ignored so far \cite{furusawa17a}.
The other models are systematically formulated to drop the washout of shell effect
or the shell effect itself.
For the electron capture rates of individual nuclei, 
we have utilized the  tabulated data of  the individual nucleus, whenever
available, and the approximation formula 
for the nuclei with no data available.

Utilizing the trajectory
 of density, temperature 
and electron fraction in the recent simulation of core-collapse,
we have made a systematic comparison of the weak interaction rates
derived with composition in 
the different EOS models.
We show that not  only nuclei with neutron magic numbers 
 but also non-magic nuclei
 contribute to the total  electron capture rates. 
We find that the washout of shell effect reduces the electron capture rates
 by $\sim 20\%$  at low densities and increases them by $\sim 40\%$  at high densities,
while
this effect also cuts down the neutrino-nucleus scattering by $\sim$ 15$\%$.
These changes arise  from the fact that the nuclei in the vicinity of the neutron magic numbers
are reduced and other nuclei are  populated  instead. 
The improvement of   the weak rates based on the EOS  model accounting for shell washout
 would improve the supernova simulations.

We have investigated the gaps between single- 
and multi-nucleus EOS models
by comparing the approximate weak rates for the average values and the most probable nucleus in single nucleus descriptions
 and the exact one for 
the full ensemble of nuclei.
We find that the single-nucleus description based on the average nucleus underestimates  electron capture rates 
by $\sim$ 80$\%$ at the beginning of core collapse due to
 the concentration of nuclear abundance in the vicinities of the neutron magic numbers. 
We have shown that the underestimation of neutrino-nucleus scattering is at most  $\sim$5$\%$,
the  size of which depends  on the dispersion of mass number.
The weak rates for the most probable nuclei deviate more largely from the actual values than those for average ones.

In this study, we adopt the fitting formula of the electron capture rate of Eq.~(\ref{eq:appro}) 
for the heavy and/or neutron-rich nuclei,  
although it was parametrized for pf-shell nuclei which are close to the beta-stability line.
The rates for such nuclei may diverge from what is predicted by the approximation formula \cite{juodagalvis08,sullivan16}. 
In addition, the in-medium effects are not carefully considered in the calculation of weak rates. 
To obtain  the precise weak rates during stellar core-collapse,
we require the new  experiments and theoretical studies 
which are aimed  not only at magic nuclei but also at non-magic ones  at finite densities and temperatures.
Note that 
 there remain various uncertainties in  EOS models.
Shell energies  are simply assumed to be the  difference between experimental or theoretical mass data and the original liquid drop model in our EOS.
The free energy of nucleons and the bulk energy  of heavy nuclei are calculated by the  RMF with the TM1 parameter set in this study.
The employment of another theory for them may also affect the nuclear component. 
We are also currently constructing an table for weak interaction based on the new EOS 
for supernova simulations, which will
be available 
 in the public domain.
The supernova simulations with the improved weak rates and the impact of the update on the dynamics will be reported in the near future.

\begin{acknowledgments}
S.F. and H. N. are supported by Japan Society for the Promotion of
Science Postdoctoral Fellowships for Research Abroad. 
We are grateful to the Goethe Graduate Academy for the proofreading.
H. N. was partially supported at Caltech through NSF award No. TCAN AST-1333520.
Some numerical calculations were carried out on  PC cluster at Center
for Computational Astrophysics, National Astronomical Observatory of Japan.
This work is supported in part by the usage of supercomputer systems
through the Large Scale Simulation Program
(Nos. 15/16/-08,16/17-11) of High Energy Accelerator Research Organization (KEK)
and
Post-K Projects  (hp 150225, hp160071, hp160211) at K-computer, RIKEN AICS
as well as the  resources provided by
RCNP at Osaka University, YITP at Kyoto University, University of Tokyo
and JLDG.
This work was
supported by Grant-in-Aid for the Scientiﬁc Research
from the Ministry of Education, Culture, Sports, Science
and Technology (MEXT), Japan (24103006, 24244036,16H03986,15K05093, 24105008).
\end{acknowledgments}

\newpage

\bibliography{reference170128}

\newpage

\end{document}